\documentclass[twocolumn,floats,floatfix,superscriptaddress,aps,pra]{revtex4-1}
\usepackage{subfigure}
\usepackage{psfrag,graphicx}
\usepackage{bm,color}
\usepackage{latexsym}
\usepackage{epstopdf}
\usepackage[english]{babel}
\usepackage{amsmath,amssymb,amsfonts}
\usepackage{appendix}
\usepackage{bbold}
\usepackage[dvipsnames]{xcolor}

\definecolor{mygrey}{gray}{0.35}
\definecolor{myblue}{rgb}{0.2,0.2,0.8}
\definecolor{myzard}{cmyk}{0,0,0.05,0}
\definecolor{mywhite}{rgb}{1,1,1}
\definecolor{mywhite}{rgb}{1,1,1}
\definecolor{myred}{rgb}{1,0.,0.3}

\def\be{ \begin{equation}}
\def\ee{ \end{equation}}
\def\bse{  \begin{subequations}}
\def\ese{  \end{subequations}}
\def\bea#1\ea{\begin{align}#1\end{align}}

\def\bi{\begin{itemize}}
\def\ei{\end{itemize}}

\def\txt#1{\textrm{#1}}
\def\bt{\begin{tabular}}
\def\et{\end{tabular}}

\def\half{\tfrac12}

\def\3half{\tfrac32}

\def\to{\rightarrow}

\def\A{\mathbf{A}}
\def\B{\mathbf{B}}
\def\U{\mathbf{U}}
\def\C{\mathbf{C}}
\def\H{\mathbf{H}}
\def\S{\mathbf{S}}
\def\C{\mathbf{C}}
\def\D{\mathbf{D}}

\def\Q{\mathbf{Q}}
\def\P{\mathbf{P}}
\def\O{\mathbf{\Omega}}

\def\R{\mathbf{R}}
\def\Xx{\mathbf{\Xi}}

\begin{document}

\begin{abstract}
The Morris-Shore (MS) transformation is a powerful tool for decomposition of the dynamics of multistate quantum systems to a set of two-state systems and uncoupled single states.
It assumes two sets of states wherein any state in the first set can be coupled to any state in the second set but the states within each set are not coupled between themselves.
Another important condition is the degeneracy of the states in each set, although all couplings between the states from different sets can be detuned from resonance by the same detuning.
The degeneracy condition limits the application of the MS transformation in various physically interesting situations, e.g. in the presence of electric and/or magnetic fields or light shifts, which lift the degeneracy in each set of states, e.g. when these sets comprise the magnetic sublevels of levels with nonzero angular momentum.
This paper extends the MS transformation to such situations, in which the states in each of the two sets are nondegenerate.
To this end, we develop an alternative way for the derivation of Morris-Shore transformation, which can be applied to non-degenerate sets of states.
We present an approximated eigenvalue approach, by which, in the limit of small detunings from degeneracy, we are able to generate an effective Hamiltonian that is dynamically equivalent to the non-degenerate Hamiltonian.
The effective Hamiltonian can be mapped to the Morris-Shore basis with a two-step similarity transformation.
After the derivation of the general framework, we demonstrate the application of this technique to the popular $\Lambda$ three-state system, and the four-state tripod, double-$\Lambda$ and diamond systems.
In all of these systems, our formalism allows us to reduce their quantum dynamics to simpler two-state systems even in the presence of various detunings, e.g. generated by external fields of frequency drifts.
\end{abstract}

\author{K. N. Zlatanov}
\affiliation{Department of Physics, Sofia University, James Bourchier 5 blvd, 1164 Sofia, Bulgaria}
\affiliation{Institute of Solid State Physics, Bulgarian Academy of Sciences, Tsarigradsko chaussée 72, 1784 Sofia, Bulgaria}
\author{G. S. Vasilev}
\affiliation{Department of Physics, Sofia University, James Bourchier 5 blvd, 1164 Sofia, Bulgaria}
\author{N. V. Vitanov}
\affiliation{Department of Physics, Sofia University, James Bourchier 5 blvd, 1164 Sofia, Bulgaria}
\title{Morris-Shore transformation for non-degenerate systems}
\date{\today }
\maketitle


\section{Introduction}

Coherent control of quantum systems is one of the corner stones of contemporary quantum physics \citep{Allen1975, Shore1990}.
Most systems which have been well studied and for which analytical solutions \cite{Rabi1937, Landau1932, Zener1932, Stuckelberg1932, Majorana1932, Rosen1932} exist consist of only two or three quantum states \cite{Vitanov2017}.
In order to make sure that only two or three states are involved in the dynamics the physical system has to be carefully isolated in order to prevent interferences from nearby states with similar energy, which is not always easy to achieve.
One of the ways to isolate a system is by large energy separation between ground and excited states.
This separation renders the manifolds of ground and excited states to a single pair by largely detuning all other transitions, which diminishes their excitation probability.
On the other hand, if the energy separation within a manifold is much smaller than the energy of the coupling field,
as is often the case when using angular momentum states or ultrashort laser pulses, such strategy is unreliable.
Even with other state isolation techniques, such as light polarization, chirping or light induced energy shifts \cite{Rangelov2005,Warring2013,Johanning2009}, additional states often have to be included which adds an extensive complexity to the system.

Multistate systems, by themselves, have many more degrees of freedom and allow the understanding of more complicated intriguing effects.
For example, analytical multilevel solutions are essential for the most famous quantum computation algorithms \cite{Nielsen2000} whose building blocks are many-qubits coherent superposition states.
Multistate systems play an important role in effects like dark-state polaritons, electromagnetically induced transparency \cite{Zimmer2008,Finkelstein2019,Appel2006}, and multistate population transfer \cite{Pillet1993} and atom optics \cite{Goldner1994,Weitz1994prl,Featonby1998,Theuer1998}, to name just a few.

Multistate systems are far more difficult to treat than two- and three-state systems as they are described by differential equations of prohibitively high order, unless they can be reduced to simpler systems \cite{Shore2013}.
One of the most prominent techniques for such reduction is the Morris-Shore (MS) transformation \cite{Morris1983}.
It is a transformation of the basis vectors in Hilbert space, which reduces two sets of degenerate states to a number of independent two-state systems and residue "dark states", uncoupled from the interaction.
The MS transformation has further been generalized to three sets of degenerate states \cite{Rangelov2006} that can reduce the dynamics to independent three-state systems.

The mathematical description of the MS transformation requires the derivation of eigenvalues and eigenvectors of an hermitian matrix, which is not a particularly hard task.
There are however a few restrictions on the MS transformation, namely all interactions must share the same time dependence, and also all interactions must be resonant, or equally detuned from the transition frequencies.
The last condition implies degeneracy among the states in each of the two sets.
While the restriction of same time dependence for the couplings can be met with careful alignment, the condition of degeneracy restricts the applicability of the MS transformation.
Indeed, the degeneracy can easily be lifted in the presence of electric and/or magnetic fields or light shifts, for example, when these sets comprise the magnetic sublevels of levels with nonzero angular momentum.

In order to remove this limitation, in this paper we propose an extension of the MS transformation to nondegenerate sets of states.
To this end, we propose a new method of obtaining the MS transformation, which allows us to transform a non-degenerate Hamiltonian to a set of independent two-state systems.
Essentially we obtain an effective Hamiltonian, which is dynamically equivalent to the non-degenerate Hamiltonian, but whose eigenvalues are much simpler.
We further obtain the non-degenerate MS transformation by mapping the diagonalized effective Hamiltonian to the MS basis with a similarity transformation, which can be obtained by the degenerate MS Hamiltonian.
We apply our technique to the $\Lambda$ configuration \cite{Bergmann1998,Torosov2012} as it's the simplest system to which many other problems are reduced.
Often, an additional ground state participates in the interaction of the $\Lambda$ system and for that matter we also investigate the tripod system \cite{Moller2007}, also the later has importance on its own \cite{Kumar2013}.
Finally we apply our results to the double $\Lambda$ system since it is of significant interest for lasing without inversion \cite{Kocharovskaya1990,Karawajczyk1992}, non-linear optics \cite{Korsunsky1999}, EIT \cite{Liu2017} and other coherent excitation effects \cite{Hamid2019}.

This paper is organized as follows. In Section \ref{Sec:Standard MS} we introduce the standard MS transformation for degenerate systems. In Section \ref{Sec:NDeg MS transform} we describe the main idea of the effective Hamiltonian and how to find its MS transformation. Further we illustrate these
concepts with some common systems in Section \ref{Sec:Examples}. Finally we conclude our findings in Section \ref{Sec:Discussion}.



\section{Degenerate Morris-Shore transformation\label{Sec:Standard MS}}

The standard Morris-Shore transformation considers a system of $g$ degenerate
ground states coupled to $e$ degenerate excited states, as described by
the time dependent Schr\"{o}dinger equation in the usual rotating-wave approximation (RWA),
\begin{equation}
i\hbar \frac{d}{dt}\mathbf{C}(t)=\mathbf{H}(t)\mathbf{C}(t).  \label{Sch}
\end{equation}%
The Hamiltonian is a block matrix given by%
\begin{equation}
\mathbf{H}=\frac{1}{2}%
\begin{bmatrix}
-\mathbf{\Delta}(t)_{g\times g} & \mathbf{V}(t)_{g\times e} \\
\mathbf{V}^{\dag }(t)_{e\times g} & \mathbf{\Delta }(t)_{e\times e}%
\end{bmatrix}%
.  \label{H}
\end{equation}%
The detuning matrices in Eq.(\ref{H}) are all diagonal namely,
\be
\mathbf{\Delta}(t)=\Delta(t) \mathbf{1}.
\label{delta-mat}
\ee
The time dependent diagonal detunings are defined as the difference of the transition
frequency of the system and the frequency of the coupling field
\be
\Delta(t)=(\omega_i-\omega_j)-\omega,
\ee
where the indices $i$ and $j$ run over the excited and ground states respectively.
Due to the degeneracy of the system $\Delta(t)$ is the same for all pairs of coupled states.\\
The interaction matrix $\mathbf{V}(t)$ is $%
g\times e$-dimensional, whose elements are the couplings
between the ground and excited states,
\begin{eqnarray}
\mathbf{V}(t) &=&%
\begin{bmatrix}
V _{11} & V _{12} & \cdots & V _{1e} \\
V _{21} & V _{22} & \cdots & V _{2e} \\
\vdots & \cdots & \ddots & \vdots \\
V _{g1} & V _{g2} & \hdots & V _{ge}%
\end{bmatrix}.
\label{V}
\end{eqnarray}%

The idea behind the MS transformation, is to find a unitary matrix $\mathbf{U},$ such that

\begin{figure}[tb]
\includegraphics[width=75mm]{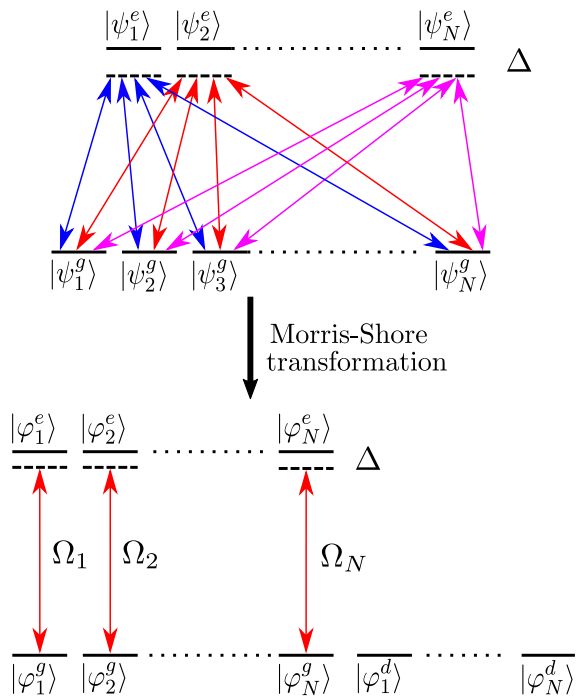}
\caption{Scheme of the Morris-Shore transformation, where a
multistate system consisting of two coupled sets of degenerate
levels is decomposed into a set of independent
two-state systems and a set of decoupled states.} \label{fig:1}
\end{figure}

\bse
\be
\C^{MS}=\mathbf{U}\C,
\ee
\be
\H^{MS}=\U\H\U^{\dagger}=\half\left[
\begin{array}{cc}
 -\mathbf{\Delta }(t) & \O(t) \\
\O^{\dagger }(t) & \mathbf{\Delta }(t)%
\end{array}%
\right],
\ee \label{MS-Basis}
\ese 
where the new interaction matrix $\mathbf{\Omega}(t)$ is now diagonal and the detuning matrix $\mathbf{\Delta}(t)$ of Eq.(\ref{delta-mat}) is
left unchanged by the transformation.
The consequence of this change of basis is that $\mathbf{H}^{MS}$ can further be
rearranged by inspection in a block-diagonal form
\be
\widetilde{\H}^{MS}=\left[
\begin{array}{cccc}
 \H^{MS}_1 & \hdots & \hdots & 0 \\
  \vdots & \ddots & \hdots & \vdots \\
   \vdots &  \hdots  & \H^{MS}_n  & \vdots\\
 	0 & \hdots & \hdots & \H^{MS}_{n+1} \\
\end{array}
\right],
\ee
such that the $n$ MS Hamiltonians are $2\times2$-dimensional and the $n+1$-st MS Hamiltonian is a diagonal matrix driving
the evolution of uncoupled spectator or "dark" MS states, as illustrated in Fig.~\ref{fig:1}.\\
The standard procedure to find the transformation matrix $\mathbf{U}$ is
to represent it in block-matrix form
\begin{equation}
\mathbf{U}=\left[
\begin{array}{cc}
\mathbf{A} & \mathbf{O} \\
\mathbf{O} & \mathbf{B}%
\end{array}%
\right] ,  \label{S0}
\end{equation}%
where $\mathbf{A}$ is a unitary $g$-dimensional square matrix and $%
\mathbf{B}$ is a unitary $e$-dimensional square matrix. We require that $\mathbf{A}$ and $\mathbf{B}$ only mix
sublevels of the ground and excited states respectively as well as diagonalize $\mathbf{V}(t)$ in a way that
\be
\O=\mathbf{AVB}^{\dagger }.  \label{Omega}
\ee%
The diagonalization of $\O(t)$ is equivalent to $\mathbf{A}$ and $\mathbf{B}$ diagonalizing the
matrices $\mathbf{VV}^{\dagger }$ and $\mathbf{V}^{\dagger }\mathbf{V}$ so that
\begin{subequations}
\label{MM}
\begin{eqnarray}
\O \O^{\dagger } &=&\mathbf{AVV}^{\dagger }\mathbf{A}^{\dagger },  \label{OO+} \\
\O^{\dagger }\O &=&\mathbf{BV}^{\dagger }\mathbf{VB}^{\dagger }.  \label{O+O}
\end{eqnarray}%
\end{subequations}
Solving Eqs.(\ref{MM}) for $\A$ and $\B$ determines $\U.$\\
The diagonal choice of $\U$ ensures that the ground and excited states in the MS basis will be, respectively superpositions purely of the ground and excited states of the original basis.\\
The MS transformation provides a powerful tool for treating multilevel systems, by simply reducing the dynamics
to a number of independent two-level systems and residue uncoupled dark states. However the requirement that the
ground and excited states are degenerate is rather strong and becomes inaccurate in the presence of various energy
shifts caused by external fields or other effects. In the next section we show a procedure that can overcome the condition of degeneracy.

\section{MS with nondegenerate(unequal) detunings}\label{Sec:NDeg MS transform}
 Whenever the energies of the ground states and the excited states are different
the detunings for all coupled states are no longer the same.
The energy shifts between different sub-levels of the ground and excited levels can be incorporated in the diagonal matrix
\begin{equation}
\mathbf{D}=
\begin{bmatrix}
\mathbf{D}_{g}(t) & \mathbf{0} \\
\mathbf{0} & \mathbf{D}_{e}(t)%
\end{bmatrix},\label{DM}
\end{equation}
whose sub-matrices
\be
\mathbf{D}_i(t)= \delta_i \mathbf{1}_{i\times i},
\label{delta1}
\ee
are also diagonal with $\delta_i$ being the frequency shift that lifts the degeneracy.
The new Hamiltonian can then be expressed as
\bea
\mathbf{H}&=\frac{1}{2}%
\begin{bmatrix}
-\mathbf{\Delta}(t)_{g\times g}+\mathbf{D}_{g}(t)_{g\times g} & \mathbf{V}(t)_{g\times e} \\
\mathbf{V}^{\dag }(t)_{e\times g} & \mathbf{\Delta }(t)_{e\times e}+\mathbf{D}_{e}(t) _{e\times e}%
\end{bmatrix}\notag \\%
&=\H_0+\mathbf{D},\label{Hdetuned}
\ea%
where $\H_0$ is the degenerate Hamiltonian of Eq.(\ref{H}) and $\D$ carries the energy shifts.\\
The complications arising from this non-degeneracy prevent us from using the standard MS transformation by changing basis with $\U$ since the Hamiltonian of Eq.(\ref{Hdetuned}) is transformed in the MS basis as%
\begin{eqnarray}
&\mathbf{H}^{MS}=\mathbf{U}(\mathbf{H}_{0}+\mathbf{D})\mathbf{U}^{\dagger }=\mathbf{H%
}_{0}^{MS}+\mathbf{UD}\mathbf{U}^{\dagger }\notag\\
 &=\left[
\begin{array}{cc}
-\mathbf{\Delta }(t)+\mathbf{A\mathbf{D}}_{g}(t)\mathbf{A}^{\dagger } & \O \\
\O^{\dagger } & \mathbf{\Delta }(t)+\mathbf{B\mathbf{D}}%
_{e}(t)\mathbf{B}^{\dagger }%
\end{array}%
\right].
\end{eqnarray}
The matrices of the additional term $\mathbf{UD}(t)\mathbf{U}^{\dagger }$, must also be diagonal for the standard MS transformation to remain valid.
For the matrix representation of $\U$ in Eq.(\ref{S0}) this is not the case since in general $\mathbf{A}\mathbf{D}_{g}(t)\mathbf{A}^{\dagger }\neq $ $%
\mathbf{\mathbf{D}}_{g}(t)$ and $\mathbf{B\mathbf{D}}_{e}(t)\mathbf{B}^{\dagger }\neq $ $\mathbf{\mathbf{D}}_{e}(t).$ Although the off-diagonal terms of $\U\D\U^{\dagger}$ are of the order of the degeneracy, they can not simply be ignored, since they introduce couplings among the 
otherwise independent Hamiltonians and more importantly among the potential dark states.
This simply means that the MS transformation has to be found by a different procedure than the one used in Section \ref{Sec:Standard MS}.
%
%
\\An alternative way to achieve the MS transformation is to first diagonalize the Hamiltonian, and then carry a second transformation that will generate the MS Hamiltonian as,
\be
\H_0\xrightarrow{\S\H_0\S^{\dagger}}\mathbf{\Xi}\xrightarrow{\P \mathbf{\Xi}\P^{\dagger}}\H^{MS}_0,
\ee
 where 	
\be
\mathbf{\Xi}=\text{diag}(\chi_1,\chi_2,...,\chi_n)
\ee
is a diagonal matrix of the eigenvalues of $\H_0$ and the transformation matrices $\S$ and $\P$ satisfy
\bea
&\S\H_0\S^{\dagger}=\mathbf{\Xi}=\P^{\dagger}\H^{MS}_0\P.
\ea
 As it is evident, $\S$ is defined as the matrix which diagonalizes $\H_0$ and $\P$ as the matrix which diagonalizes $\H_0^{MS}.$ Combined together $\S$ and  $\P$ achieve
\bea
\P\S\H_0 \S^{\dagger}\P^{\dagger}=\U\H_0\U^{\dagger}=\H^{MS}_0,
\ea
with 
\be
\U=\P\S. \label{UPS}
\ee
In general the two-step approach preserves the structure of diagonal sub-matrices in the MS Hamiltonian, generated by a single step transformation. Since this structure is the same for both, single and two-step transformations it may also need to be further rearranged in block diagonal form by a consequent similarity transformation.

 We now want to find an effective Hamiltonian which is dynamically equivalent approximation of the non-degenerate Hamiltonian and can also be transformed to the MS basis by Eq.(\ref{UPS}). This will be the case if the effective Hamiltonian has approximately the same eigenvalues as Eq.(\ref{Hdetuned}).
In order to utilize the two-step approach we choose the effective Hamiltonian as
\be
\H_{eff}=\S^{\dagger} \Q\S \H_0,\label{H_eff}
\ee
where the matrix $ \Q$ is to be determined.
 The specific form of $\H_{eff}$ becomes clear once we transform it to the MS basis, which reads
\be
\U\H_{eff}\U^{\dagger}=\P\S\S^{\dagger}\Q\S\H_0\S^{\dagger}\P^{\dagger}=\P\Q\mathbf{\Xi}\P^{\dagger}=\H^{MS}.
\label{Heff_transform}
\ee
 The procedure by which we find the $\P$ matrix ensures that its non-zero elements are at the correct positions so that upon similarity transformation with $\P$ the transformed matrix has diagonal sub-matrices.
In order to ensure that Eq.(\ref{Heff_transform}) is the proper MS transformation of the non-degenerate Hamiltonian we have to set $\Q$ such that
\be
\S\H_{eff}\S^{\dagger}= \Q\mathbf{\Xi}=\mathbf{W}=\R\H\R^{\dagger}
\ee
holds. Here
\be
\mathbf{W}=\text{diag}(\varepsilon_1,\varepsilon_2,...,\varepsilon_i)
\ee
is the matrix of the eigenvalues of the non-degenerate Hamiltonian of Eq.(\ref{Hdetuned}) and $\R$ is composed of its eigenvectors. The matrices $\R$ and $\mathbf{W}$ can be found by any standard diagonalization procedure.
In order to find the matrix $ \Q$ we note that the non-degenerate eigenvalues can be expressed as a series expansion in terms of the energy shifts,
\be
\varepsilon_i=\sum_{k=0}^\infty \frac{\delta_i^k}{k!}\frac{d^k\varepsilon_{i}}{d\delta_i^k}|_{\delta_i=0},
\ee
where $\delta_i$ is the appropriate energy shift that corresponds to the $i$-th eigenvalue.
Whenever these shifts are small enough we can drop the higher-order terms and only keep the linear expansion
\be
\varepsilon_i\approx\chi_i+\delta_i \kappa_{i},
\label{EV_approx}
\ee
where
\be
\kappa_i=\frac{d\varepsilon_i}{d\delta_i}|_{\delta_i=0}
\ee
is a function of the control parameters of the Hamiltonian, that is independent of $\delta_i$. Whenever more than one energy shift is involved in a specific eigenvalue the appropriate vector form of Eq.(\ref{EV_approx}) should be used. \\
 The simplest choice for the matrix $ \Q$ will be a diagonal form, whose $i$-th diagonal element reads
\be
 \text{Q}_{i}=1+\frac{\delta_i\kappa_i}{\chi_i},\label{Belements}
\ee
since this choice yields $\varepsilon_i\approx\text{Q}_{i}\chi_i$.

Whenever an eigenvalue is a zero, it should instead be set to $\chi_{0}=\displaystyle{\lim_{ p \to 0}p},$ as well as the corresponding element in $ \Q$ reads
\be
\text{Q}_{0}= \lim_{ p \to 0}\left(1+\frac{\delta_0\kappa_0}{p}\right).\label{Bzero}
\ee
In this way the eigenvalue evaluates to
\be
\varepsilon_0\approx\text{Q}_0\chi_0=\delta_0\kappa_0.
\ee

 We note that the non-degenerate MS Hamiltonian can be obtained by acting with $\U$ on the effective Hamiltonian or with $\P$ on $\Q\mathbf{\Xi}$ as evident from Eq.(\ref{Heff_transform}). 
Later on in the text we will use the $\P$ matrix on the approximated diagonal form of $\H$ to generate the MS transformation.\\
To summarize, our approach has the following sequence of steps. First we find the MS transformation for the degenerate Hamiltonian which gives the matrix $\U.$ Then we diagonalize the degenerate MS Hamiltonian to find the map between the diagonal form and the MS basis, which yields the $\P$ matrix and by Eq.(\ref{UPS}) the $\S$ matrix. The third step is to diagonalize the non-degenerate Hamiltonian and express its eigenvalue matrix as
\be
\mathbf{W}= \Q\Xi,
\ee
by which we find $\mathbf{W}$ and  construct the $ \Q$ matrix.
Finally the non-degenerate MS Hamiltonian is obtained by Eq.(\ref{Heff_transform}).

In the next section we illustrate this  two-step approach to the MS transformation with some common systems of high practical significance, namely the $\Lambda$, the tripod, and the double $\Lambda$.

\section{Specific examples} \label{Sec:Examples}

\subsection{Lambda system}

\begin{figure}[tb]
\includegraphics[width=0.70\columnwidth]{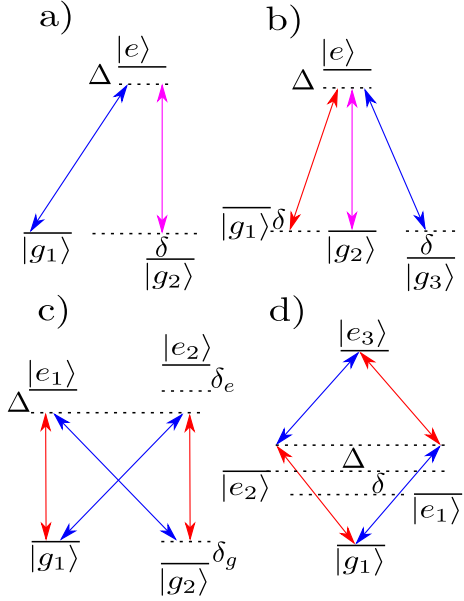}
\caption{Non-degenerate a)$\Lambda$ system, b) tripod system, c) double $\Lambda$ system and d) diamond system.}
    \label{fig:systems}
\end{figure}

The simplest case we consider is a $\Lambda$ system, whose final state has a different energy from the initial state as shown in Fig.~\ref{fig:systems}(a).

In order to focus on the MS transformation and to simplify our calculations we will only consider resonant excitation where $\Delta(t)=0$, although our results can be applied for non-resonant excitation as well.
The non-degenerate Hamiltonian reads
\be
\H_{\Lambda}=\frac{1}{2}\left[
\begin{array}{ccc}
 0 & 0 & \Omega _s \\
 0 & 2\delta  & \Omega _p \\
 \Omega_s & \Omega_p & 0  \\
\end{array}
\right],\label{H_Lambda}
\ee
where for the sake of simplicity we have assumed the Rabi frequencies to be real.
After we apply the procedure from Section \ref{Sec:Standard MS} we find the transformation matrix for the degenerate system to be
\be
\U=\frac{1}{\Omega_{rms}}\left[
\begin{array}{ccc}
 -\Omega _p & \Omega _s & 0 \\
 \Omega _s & \Omega _p & 0 \\
 0 & 0 & \Omega_{rms} \\
\end{array}
\right],
\ee
where we have introduced the root mean square Rabi frequency
\be
\Omega_{rms}=\sqrt{\Omega _p^2+\Omega _s^2}.
\ee
The degenerate MS Hamiltonian is then
\be
\H_{\Lambda|\delta=0}^{MS}=\frac{1}{2}\left[
\begin{array}{ccc}
 0 & 0 & 0 \\
 0 & 0 &  \Omega_{rms} \\
 0 & \Omega_{rms}& 0  \\
\end{array}
\right].\label{HMS_Lambda}
\ee

Furthermore, its diagonalization reads
\be
\P^{\dagger}\H^{MS}_{\Lambda|\delta=0}\P=\Xx=\frac{1}{2}\left[
\begin{array}{ccc}
 0 & 0 & 0 \\
 0 & - \Omega_{rms} & 0 \\
 0 & 0 & \Omega _{rms} \\
\end{array}
\right],
\ee
with the matrix $\P$ given as
\be
\P=\left[
\begin{array}{ccc}
 1 & 0 & 0 \\
 0 & -\frac{1}{\sqrt{2}} & \frac{1}{\sqrt{2}} \\
 0 & \frac{1}{\sqrt{2}} & \frac{1}{\sqrt{2}} \\
\end{array}
\right].
\ee
The eigenvalues of the non-degenerate Hamiltonian are too cumbersome to be presented here so instead we directly present the result for the approximated eigenvalues. From Eq.(\ref{EV_approx}), Eq.(\ref{Belements}) and Eq.(\ref{Bzero}) we find them to be
\be
 \Q\Xx=\left[
\begin{array}{ccc}
 \frac{\delta  \Omega _s^2}{\Omega _{rms}^2} & 0 & 0 \\
 0 & \frac{\delta  \Omega _p^2}{2 \Omega _{rms}^2}-\frac{\Omega _{rms}}{2} & 0 \\
 0 & 0 & \frac{\delta  \Omega _p^2}{2 \Omega _{rms}^2}+\frac{\Omega _{rms}}{2} \\
\end{array}
\right].\label{BX}
\ee
\begin{figure}[tb]
\includegraphics[width=75mm]{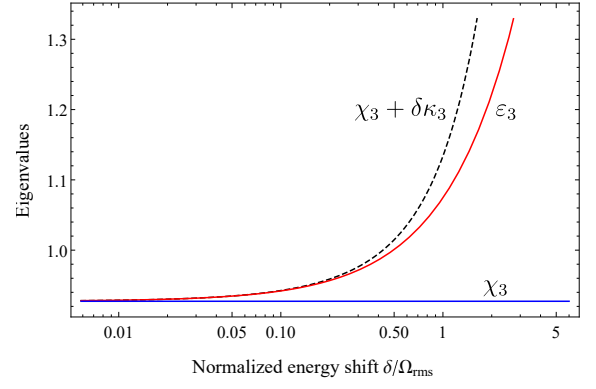}
\caption{Comparison between the third  approximated eigenvalue of Eq.(\ref{BX}) and its corresponding exact eigenvalue of Eq.(\ref{H_Lambda}).
The parameters have been set to $\Omega_s=1.25$ $T^{-1}$, $\Omega_p=1.37$ $T^{-1},$ and $\Omega_{rms}=1.85$ $T^{-1},$ where $T$ is the pulse duration for the excitation.} \label{fig:5}
\end{figure}

In order to estimate how good our approximation is in Fig.~\ref{fig:5} we have compared the bottom corner  approximated eigenvalue of Eq.(\ref{BX}) with its corresponding eigenvalue of Eq.(\ref{H_Lambda}). As evident from the figure the approximation holds quite well as long as the ratio $\delta/\Omega_{rms}$ is of  $\mathcal{O}(10^{-2})$ order.
The MS Hamiltonian (\ref{Heff_transform}) for the non-degenerate $\Lambda$ system reads
\be
\H^{MS}_{\Lambda}=\left[
\begin{array}{ccc}
 \frac{\delta  \Omega _s^2}{\Omega _{rms}^2} & 0 & 0 \\
 0 & \frac{\delta  \Omega _p^2}{2 \Omega _{rms}^2} & \frac{\Omega _{rms}}{2} \\
 0 & \frac{\Omega _{rms}}{2} & \frac{\delta  \Omega _p^2}{2 \Omega _{rms}^2} \\
\end{array}
\right].
\ee
The $\Lambda$ system provides a dark state uncoupled from the evolution of the other two. Often a third ground state coupled to the excited state is added forming the tripod system, which we explore next.

\subsection{Tripod system}
The tripod system is illustrated in Fig.~\ref{fig:systems}(b). We assume that the middle coupling is resonant, while the "left" and "right" couplings are detuned with $\mp \delta$ respectively due to the lifted degeneracy. The Hamiltonian of the system reads,
\be
\H_{\rm{T}}=\frac{1}{2}\left[
\begin{array}{cccc}
 -\delta  & 0 & 0 & \Omega _p \\
 0 & 0 & 0 & \Omega _s \\
 0 & 0 & \delta  & \Omega _c \\
 \Omega _p & \Omega _s & \Omega _c & 0 \\
\end{array}
\right]\label{H_tripod}.
\ee
The degenerate MS transformation is carried out by
\be
\U=\left[
\begin{array}{cccc}
 \frac{\Omega _p \Omega _s}{\Omega _{rms} \sqrt{\Omega _c^2+\Omega _p^2}} &
   -\frac{\sqrt{\Omega _c^2+\Omega _p^2}}{\Omega _{rms}} & \frac{\Omega _c \Omega
   _s}{\Omega _{rms} \sqrt{\Omega _c^2+\Omega _p^2}} & 0 \\
 \frac{\Omega _c}{\sqrt{\Omega _c^2+\Omega _p^2}} & 0 & -\frac{\Omega _p}{\sqrt{\Omega
   _c^2+\Omega _p^2}} & 0 \\
 \frac{\Omega _p}{\Omega _{rms}} & \frac{\Omega _s}{\Omega _{rms}} &
   \frac{\Omega _c}{\Omega _{rms}} & 0 \\
 0 & 0 & 0 & 1 \\
\end{array}
\right],\label{U-tripod}
\ee
where
\be
\Omega_{rms}=\sqrt{\Omega_p^2+\Omega_s^2+\Omega_c^2}.
\ee

Then the standard MS Hamiltonian is
\be
\H^{MS}_{\rm{T}|\delta=0}=\frac{1}{2}\left[
\begin{array}{cccc}
 0 & 0 & 0 & 0 \\
 0 & 0 & 0 & 0 \\
 0 & 0 & 0 & \Omega _{rms} \\
 0 & 0 & \Omega _{rms} & 0 \\
\end{array}
\right].\label{HMS_Tripod}
\ee
 Diagonalizing it with the matrix
\be
\P=\left[
\begin{array}{cccc}
 1 & 0 & 0 & 0 \\
 0 & 1 & 0 & 0 \\
 0 & 0 & \frac{1}{\sqrt{2}} & \frac{1}{\sqrt{2}} \\
 0 & 0 & -\frac{1}{\sqrt{2}} & \frac{1}{\sqrt{2}} \\
\end{array}
\right]
\ee
leaves us with
\be
\P^{\dagger}\H^{MS}_{\rm{T}|\delta=0}\P=\frac{1}{2}
\left[
\begin{array}{cccc}
 0 & 0 & 0 & 0 \\
 0 & 0 & 0 & 0 \\
 0 & 0 & -\Omega _{rms} & 0 \\
 0 & 0 & 0 & \Omega _{rms} \\
\end{array}
\right].
\ee
The approximated eigenvalue matrix reads
\be
 \Q \Xx=\left[
\begin{array}{cccc}
 \tilde{\Delta}_1 & 0 & 0 & 0 \\
 0 & \tilde{\Delta}_2 & 0 & 0 \\
 0 & 0 &\frac{\delta  \tilde{\Omega}_-}{4
   \Omega _{rms}^2}-\frac{\Omega _{rms}}{2} & 0 \\
 0 & 0 & 0 &\frac{\delta  \tilde{\Omega}_-}{4
   \Omega _{rms}^2}+\frac{\Omega _{rms}}{2} \\
\end{array}
\right],\label{DiagHTripod}
\ee
where we have substituted
\bse
\bea
&\tilde{\Omega}_+=\Omega
   _p^2+\Omega _c^2,\\
&\tilde{\Omega}_-=\Omega _p^2-\Omega _c^2 ,  \\
&\tilde{\Delta}_1=-\frac{\delta  \left(\tilde{\Omega}_--\sqrt{4 \Omega _s^4+4 \tilde{\Omega}_+ \Omega _s^2+\tilde{\Omega}_-{}^2}\right)}{4
   \Omega _{rms}^2},\\
&\tilde{\Delta}_2= -\frac{\delta  \left(\tilde{\Omega}_-+\sqrt{4 \Omega _s^4+4\tilde{\Omega}_+ \Omega _s^2+\tilde{\Omega}_-{}^2}\right)}{4
   \Omega _{rms}^2}. \label{DT-2}
\ea
\ese
Finally the MS transformation of Eq.(\ref{H_tripod}) reads

\be
\H^{MS}_{\rm{T}}=\left[
\begin{array}{cccc}
 \tilde{\Delta}_1 & 0 & 0 & 0 \\
 0 & \tilde{\Delta}_2 & 0 & 0 \\
 0 & 0 & \frac{\delta  \tilde{\Omega} _-}{4 \Omega _{rms}^2} &
   \frac{\Omega _{rms}}{2} \\
 0 & 0 & \frac{\Omega _{rms}}{2} & \frac{\delta  \tilde{\Omega} _-}{4 \Omega _{rms}^2} \\
\end{array}
\right].
\ee
Both the $\Lambda$ and the tripod systems have a dark state(s), whose evolution is uncoupled from the rest of the system. For applications in quantum information for example, these are the states of interest, since they can sustain superposition for considerable times. The effect of non-degeneracy leaves a non-zero eigenvalue which contributes a global phase shift to the evolution of the system. In a degenerate system the superposition will have a global phase which is beyond experimental control since in the MS basis it is a function of the Rabi frequencies and the detuning. Thus changing them will change not only the global phase but the superposition as well. This comes as a consequence of the zero eigenvalue of the dark state. However if we instead have a non-zero eigenvalue, whose phase contribution depends on a parameter, that doesn't affect the superposition, we can control the global phase. For example a close look at Eq.(\ref{U-tripod}) reveals that the second MS state will be a superposition of the first and third state in the original basis, that is independent of $\Omega_s$. During the evolution of the system, however the state remains unchanged besides a phase shift proportional to the detuning element of Eq.(\ref{DT-2}). This phase shift depends on $\Omega_s$ thus giving a control parameter that leaves the superposition intact, while changing the global phase.

\subsection{Double $\Lambda$\label{Sec:X-system}}
The final example we discuss involves multiple excited states. The simplest case is the double $\Lambda$ shown in Fig.~\ref{fig:systems}(c), which consists of two ground and two excited states which are non-degenerate. This systems can also be represented in a diamond configuration, given in Fig.~\ref{fig:systems}(d), as the two are similar. In order to simplify the problem we assume that the direct and cross couplings between the ground states and the excited states are equal,
\begin{subequations}
\bea
\Omega_{11}=\Omega_{22}=\Omega_d,\\
\Omega_{12}=\Omega_{21}=\Omega_c,
\ea
\end{subequations}
as is the case for $J=1/2 \rightarrow J=1/2$ transitions.

The Hamiltonian reads
\bea
\label{Hx}
\H_{\txt{2}\Lambda} &=\half\left[
\begin{array}{cccc}
 0 & 0 & \Omega _d & \Omega _c \\
 0 & -\delta_g  & \Omega _c & \Omega _d \\
  \Omega _d & \Omega _c & 0& 0 \\
   \Omega _c & \Omega _d & 0 & \delta_e \\
\end{array}
\right].
\ea
Following the procedures from Section \ref{Sec:Standard MS} we find the MS Hamiltonian to be
\be
\label{HMSX}
\H^{MS}_{\txt{2}\Lambda |\delta_i=0}=\frac{1}{2}\left[
\begin{array}{cccc}
 0 & 0 & \Omega _- & 0 \\
 0 & 0 & 0 & -\Omega _+ \\
 \Omega _- & 0 & 0 & 0 \\
 0 & -\Omega _+ & 0 & 0 \\
\end{array}
\right],
\ee
where we have taken the shorthand notation
\bse
\bea
&\Omega_{+}=\Omega_c+\Omega_d,\\
&\Omega_{-}=\Omega_c-\Omega_d.
\ea
\ese
The transformation matrix to the MS basis is then
\be
\U=\left[
\begin{array}{cccc}
 \frac{1}{\sqrt{2}} & -\frac{1}{\sqrt{2}} & 0 & 0 \\
 -\frac{1}{\sqrt{2}} & -\frac{1}{\sqrt{2}} & 0 & 0 \\
 0 & 0 & -\frac{1}{\sqrt{2}} & \frac{1}{\sqrt{2}} \\
 0 & 0 & \frac{1}{\sqrt{2}} & \frac{1}{\sqrt{2}} \\
\end{array}
\right].
\ee
The next step is to find the map between the MS Hamiltonian and the diagonal form, which reads
\be
\P^{\dagger}\H^{MS}_{\txt{2}\Lambda |\delta_{e,g}=0}\P=\frac{1}{2}\left[
\begin{array}{cccc}
 -\Omega _{+}& 0 & 0 & 0 \\
 0 & \Omega _{-}& 0 & 0 \\
 0 & 0 & -\Omega _{-}& 0 \\
 0 & 0 & 0 & \Omega _{+}\\
\end{array}
\right],
\ee
with diagonalization matrix
\bea
\P=\left[
\begin{array}{cccc}
 0 & \frac{1}{\sqrt{2}} & -\frac{1}{\sqrt{2}} & 0 \\
 \frac{1}{\sqrt{2}} & 0 & 0 & -\frac{1}{\sqrt{2}} \\
 0 & \frac{1}{\sqrt{2}} & \frac{1}{\sqrt{2}} & 0 \\
 \frac{1}{\sqrt{2}} & 0 & 0 & \frac{1}{\sqrt{2}} \\
\end{array}
\right].
\ea
From Eq.(\ref{EV_approx}) and Eq.(\ref{Hx}) we find the approximated eigenvalue matrix to be
\be
 \Q\Xx=\left[
\begin{array}{cccc}
 \delta_--\Omega _+& 0 & 0 & 0 \\
 0 & \delta_-+\Omega _-& 0 & 0 \\
 0 & 0 & \delta_--\Omega _- & 0 \\
 0 & 0 & 0 & \delta_-+\Omega _+\\
\end{array}
\right],
\ee
where
\be
\delta_-=\frac{\delta_e -\delta_g }{4}.
\ee
The MS Hamiltonian is then
\be
\H^{MS}_{\txt{2}\Lambda}=\frac{1}{2}\left[
\begin{array}{cccc}
 \delta_- & 0 &  \Omega _- & 0 \\
 0 & \delta_- & 0 & - \Omega _+ \\
  \Omega _- & 0 &\delta_- & 0 \\
 0 & -\Omega _+ & 0 & \delta_- \\
\end{array}
\right],
\ee
which can be transformed to a block-diagonal form with the permutation matrix
\be
\mathbf{\pi}=\left[
\begin{array}{cccc}
 1 & 0 & 0 & 0 \\
 0 & 0 & 1 & 0 \\
 0 & 1 & 0 & 0 \\
 0 & 0 & 0 & 1 \\
\end{array}
\right],
\ee
that finally leaves us with
\be
\mathbf{\pi}\H^{MS}_{\txt{2}\Lambda}\mathbf{\pi}^{\txt{T}}=\frac{1}{2}\left[
\begin{array}{cccc}
 \delta_- & \Omega _- & 0 & 0 \\
 \Omega _- & \delta_- & 0 & 0 \\
 0 & 0 & \delta_- & -\Omega _+ \\
 0 & 0 & -\Omega _+ & \delta_- \\
\end{array}
\right].\label{HMS-Xsystem}
\ee
 The clear benefit of the MS transformation here is the simple two-state picture of the excitation dynamics. Moreover when a number of
states, as well as control parameters are involved, the dynamics of the system is not at all obvious. For example in our current case if
we set $\Omega_c=\Omega_d$ we will have no coupling in the upper block of Eq.(\ref{HMS-Xsystem}). Furthermore if we initially set all the population in the MS states corresponding to that block, there will be no excitation of any MS state. This conclusion can't be drawn from Eq.(\ref{Hx}) directly, but comes easily in the MS basis.

\section{Discussion and Conclusion}\label{Sec:Discussion}

In this paper, we explored the extension of the MS transformation to sets of non-degenerate states.
For this purpose, we developed an approach based on  two-step transformation that carries an effective Hamiltonian to the MS basis.

A key point in our analysis and derivation of the effective Hamiltonian of Eq.\eqref{H_eff} is that its eigenvalues match those of the non-degenerate Hamiltonian with high accuracy.
We choose the specific form of $\H_{eff}$ in order to keep the validity of the MS transformation, that is originally derived for the degenerate Hamiltonian, as well as to ensure its dynamical equivalence to the non-degenerate Hamiltonian. 
We assume that the energy shifts just slightly perturb the non-degenerate eigenvalues, which justifies the linear approximation in Eq. \eqref{EV_approx}.
In addition, the approximation holds strongly for $\delta / \Omega_{rms}\sim \mathcal{O}(10^{-2}),$ which describes well optical transitions where the excitation frequency is greater than a THz and the frequency shift among magnetic sublevels, for example, is of the order of MHz. 
If this is not the case, higher order terms have to be included into the eigenvalue approximation of the effective Hamiltonian.

We illustrated our concept explicitly with four popular systems, namely the $\Lambda$, tripod, double-$\Lambda$ and diamond systems, which have numerous applications in a variety of physical situation.
%
%
Further applications of our results might be expected in the calculation of state fidelities. 
When an experimental graph of a readout measurement is compared with a theoretical prediction that assumes degeneracy, a discrepancy is to be expected. The cause is due to a variety of factors, that are often prescribed to experimental imperfections rather than inaccuracy of the theoretical model.
We expect that our model will account for such discrepancies between laboratory measured fidelity and theoretical predictions due to negligence of non-degeneracy. The magnitude of this effect remains to be investigated.


\acknowledgments 
KNZ acknowledges support from the project MSPLICS - P.Beron Grant from The Bulgarian National Science Fund (BNSF), 
GSV acknowledges support from the project QUANTNET - European Reintegration Grant (ERG) - PERG07-GA-2010-268432.

\end{document}